\documentclass[twocolumn,showpacs,superscriptaddress]{revtex4}  %use this for length estimate
\usepackage{graphicx}% Include figure files
\usepackage{dcolumn}% Align table columns on decimal point
\usepackage{bm}% bold math

\begin{document}
%\preprint{APS/123-QED}

\title{An Optical Lattice on an Atom Chip}
\date{\today}
\pacs{}

\author{D. Gallego}
 \affiliation{Physikalisches Institut, Universit\"at Heidelberg, 69120 Heidelberg, Germany}
%\author{St. Wildermuth}
% \affiliation{Physikalisches Institut, Universit\"at Heidelberg, 69120 Heidelberg, Germany}
\author{S. Hofferberth}
 \affiliation{Physikalisches Institut, Universit\"at Heidelberg, 69120 Heidelberg, Germany}
 \affiliation{Department of Physics, Harvard University, Cambridge, MA 02138, USA}
\author{T. Schumm}
 \affiliation{Physikalisches Institut, Universit\"at Heidelberg, 69120 Heidelberg, Germany}
 \affiliation{Atominstitut der \"Osterreichischen Universit\"aten, TU-Wien, 1020 Vienna, Austria}
 %\email{schumm@atomchip.org}
 \author{P. Kr\"uger}
 \affiliation{Physikalisches Institut, Universit\"at Heidelberg, 69120 Heidelberg, Germany}
 \affiliation{Midlands Ultracold Atom Research Centre, The University of Nottingham, Nottingham NG7 2RD, United Kingdom}
\author{J. Schmiedmayer}
%\email[]{Schmiedmayer@AtomChip.org}
 \affiliation{Physikalisches Institut, Universit\"at Heidelberg, 69120 Heidelberg, Germany}
 \affiliation{Atominstitut der \"Osterreichischen Universit\"aten, TU-Wien, 1020 Vienna, Austria}

\begin{abstract}
%Atom chip microtraps can be combined with optical lattice
%potentials when far detuned laser beams are reflected off the high
%quality gold layers on the chip surface. Using an atom chip based
%1d standing wave optical lattice we load atoms from a magnetic
%microtrap into a purely optical 2d trap. In a second experiment we
%demonstrate coherent Bloch oscillations in the 1d lattice.
Optical dipole traps and atom chips are two very powerful tools for the quantum manipulation of neutral atoms. We demonstrate that both methods can be combined by creating an optical lattice potential on an atom chip. A red-detuned laser beam is retro-reflected using the atom chip surface as a high-quality mirror, generating a vertical array of purely optical oblate traps. We load thermal atoms from the chip into the lattice and observe cooling into the two-dimensional regime where the thermal energy is smaller than a quantum of transverse excitation. Using a chip-generated Bose-Einstein condensate, we demonstrate coherent Bloch oscillations in the lattice.
\end{abstract}

\maketitle

Engineering quantum states on a microscale has been a long standing goal in atomic physics and quantum optics. Atom chips \cite{Folman2002,Fortagh2007} provide a robust technological basis for the manipulation of neutral atoms, e.g. coherent manipulation of internal (spin) \cite{Treutlein2004} and external (motional) \cite{Schumm2005b} states has been demonstrated. So far, atom chip traps mainly rely on the interaction of neutral atoms with static (or slowly varying) magnetic \cite{Folman2002,Fortagh2007} or electric fields \cite{Krueger2003,electricchip}. This imposes a restriction to magnetically trappable, weak-field seeking atomic states, which represents a severe limitation. For example, using Feshbach resonances \cite{FeshbachReview} to modify the atom-atom interactions becomes very difficult.

This limitation can be overcome by employing rapidly time-varying fields. The interaction with oscillating fields can be strongly enhanced when coupling to internal states of the trapped atoms. Prime examples are optical dipole potentials \cite{Grimm2000} and radio frequency (RF) induced dressed state potentials \cite{Lesanovsky2006}. Whereas RF induced potentials allow very versatile atom manipulation \cite{Hofferberth2006}, they are still linked to magnetically trappable states. The optical dipole potential in contrast gives complete freedom in choosing the trapped state.

In this Letter we present experiments combining standard atom chip technology with optical dipole trapping. To demonstrate the feasibility of such a concept, we implement an optical lattice potential, combining strong atomic confinement and high flexibility with respect to manipulating {\em all} internal atomic states with precise local manipulation and single-site addressability.

The approach of a hybrid magnetic-optical atom chip significantly enhances the flexibility of the experimental system. As dicussed above, optical traps \cite{Grimm2000} enable trapping also of magnetic strong-field seeking states. This makes manipulation of atomic scattering properties by Feshbach resonances and creation of molecules \cite{FeshbachReview} possible. Using red-detuned lasers implements purely optical trapping close to the atom chip, freeing the magnetic fields for other tasks. Fast, robust, and spatially tailored addressing of Feshbach resonances using atom chip near fields hence becomes possible. Moreover the chip fields bring local addressability to the optical lattice as required in many schemes for quantum information processing with neutral atoms \cite{Schrader2004}. The application of static or RF magnetic fields created with the microscopic chip structures will allow local state selective manipulation.

An optical lattice potential can be implemented conveniently on atom chips where the chip surface consists of a high quality gold layer which serves as a mirror to create a standing light wave. Structuring the atom chip surface (e.g. by height variations or patterning reflecting and absorbing elements) allows to design intrinsically stable and versatile optical potentials. These can range from designed circuits to high resolution random potentials from speckles.

\begin{figure}[t]\center
   \includegraphics[angle=0,width=\columnwidth]{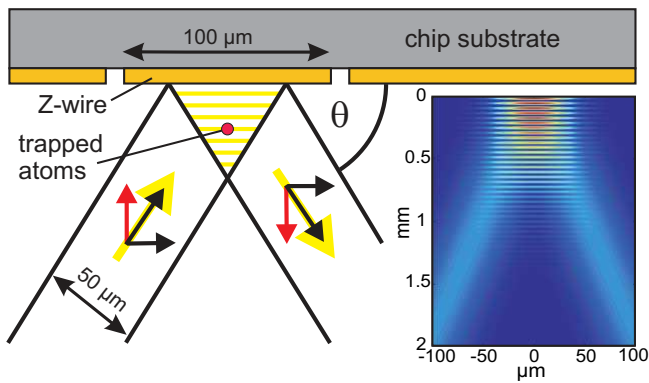}
   \caption{(\emph{color on-line}) Setup to create an optical lattice on an atom chip (not to scale, $\theta\sim 87 ^{\circ}$). A focused laser beam (50\,$\mu$m) is reflected from the central part of a 100 $\mu$m wide Z-shaped trapping wire and forms a standing light wave (inset, $\lambda$ multiplied by 100 for visualization).}
   \label{fig:OL@chip_setUp}
\end{figure}

In the experiments discussed in this Letter we use a very simple implementation of an optical lattice potential on an atom chip: A single red-detuned laser beam impinges on the chip surface at almost normal incidence. Interference with the reflected beam creates a vertical lattice of oblate (`pancake'-shaped) traps, all parallel to the chip surface (Fig.\ \ref{fig:OL@chip_setUp}). In these potentials, atoms can be confined to two dimensions (2d) analogous to electrons in a quantum well \cite{Gau1998}. The proximity of the chip surface allows to structure the 2d optical trap with the chip fields. Additional potential minima can be formed via electric fields. Magnetic field minima or maxima in the 2d plane of the optical trap will repel (attract) the atoms depending on their internal state. Multiple laser beams facilitate more sophisticated 2d or 3d potential structuring. The global arrangement of traps will in all cases be a structure of layers parallel to the reflecting chip surface.

To demonstrate the feasibility of combining optical trapping with an atom chip we performed two proof-of-principle experiments. In the first experiment we load atoms into the optical lattice and cool them into the 2d regime. In the second experiment we show that atoms can undergo coherent Bloch oscillations in the lattice close to the atom chip surface.

Our experimental apparatus is based on a hybrid macroscopic-microscopic atom chip assembly \cite{Wildermuth2004}.
It holds the macroscopic wire structures used to capture and pre-cool the atoms in the primary phase of the experiment as well as the micro structures needed for trapping and cooling the sample to quantum degeneracy. The micro structures on the atom chip are fabricated from a thermally evaporated gold layer (3.1\,$\mu$m) deposited on a $700\,\mu$m thick silicon substrate and patterned using UV-light lithography \cite{Groth2004}. Away from wire edges, the resulting mirror has a reflectivity of $>95\%$, so that an optical lattice can conveniently be formed by shining a focused laser beam ($w\sim 50 \mu$m) onto a broad (100\ $\mu$m) wire that is also used in the magnetic trapping and cooling procedure (Fig.\ \ref{fig:OL@chip_setUp}). The laser light (up to 8 mW) is derived from a laser diode, tuned to 782 nm, 2 nm above the D2 transition in $^{87}$Rb. The beam is reflected under an angle of $\theta\sim 87 ^{\circ}$ to the chip surface. The vertical distance between the trapping layers is $d=\frac{\lambda}{2 \sin \theta}= 392$ nm. The transverse size of the trap is given by the overlap of the laser beam with its reflection.

A cold sample of typically $10^5$ $^{87}$Rb atoms in the F=m$_F$=2 state just above or within the quantum degenerate regime is prepared in a magnetic microtrap following the procedure described in \cite{Wildermuth2004}. All information about the ultra-cold atomic sample is extracted from resonant absorption imaging, performed on the $\mathrm{F}=2\rightarrow \mathrm{F}'=3$ transition, using standard precision achromats with a diffraction limited resolution of $\sim 3$ $\mu$m. 

In the first demonstration experiment we load a cloud of thermal atoms ($T \sim 5 \, \mu$K) into the planes of the optical lattice, realizing purely optical trapping. This is accomplished by first adiabatically switching on the lattice and then adiabatically switching off the magnetic trap holding the atoms. Typically 10 lattice sites are then populated, a slight mismatch in the longitudianl extension of the atom cloud trapped in the magnetic and the optical trap reduces the transfer efficiency to 88 $\%$. After an adjustable holding time, the trapped atoms are released from the lattice and observed in free expansion. Experiments are performed with different lattice depths ranging from 32 $\mu K$ and $\omega_{tr}=2\pi\times 100$ kHz transverse confinement to a shallow lattice of only 4 $\mu K$ depth with $\omega_{tr}=2\pi\times 35$ kHz transverse confinement (see table \ref{Tab:2dTrap}).

\begin{table}
  \centering
  \begin{tabular}{|c|c|c|c|c|c|c|}
   \hline
   $ U_\textrm{dip}$ &$\omega_\textrm{2d}/2\pi$ & $\omega_\textrm{tr}/2\pi$ & heating & $T_\textrm{load}$  &$T_\textrm{2d, final}$& $T_\textrm{tr, final}$ \\
       \[   [$\mu$K]        &         [Hz]        &     [kHz]     &[$\mu$K/s]&   [$\mu$K]      &      [$\mu$K]      &      [$\mu$K]      \\
   \hline
    35      &     350     &     100     &   14   &    5     &     2.4    &     2.4     \\
    20      &     280     &      80     &   10   &    5     &     1.7    &     1.8     \\
    4       &     125     &      36     &    2   &    5     &     0.35   &     0.6     \\
  \hline
  \end{tabular}
  \caption{Parameters for the 2d traps in the optical lattice.}\label{Tab:2dTrap}
\end{table}

After loading, we observe a fast initial loss of atoms on the timescale of about 40 ms. Together with this initial decay we observe significant cooling of the atoms (see table \ref{Tab:2dTrap}). We hence attribute this fast loss to plain evaporation due to the limited depth of the optical trap, the high ratio of trap depth to final temperature indicates very efficient cooling. Beyond the initial decay we find a slower loss process (decay constant $\tau \sim 100-200$ ms, depending on laser intensity) at approximately constant temperature. The cooling appears to be compensated by heating processes induced for example by spontaneous photon scattering in the lattice with small detuning (calculated rates in table \ref{Tab:2dTrap}). The heating in turn translates into loss, again due to the finite trap depth.

For the shallow (low intensity) optical trap we clearly observe an anisotropic momentum distribution in TOF at the end of the first plain evaporation process. The expansion in the transverse, strongly confining direction is significantly larger than in the 2d plane ($T_\textrm{tr} = 0.6\,\mu $K compared to $T_\textrm{2d} = 0.35 \,\mu$K). This behavior indicates that the zero point motion of the trap ground state starts to dominate the momentum distribution. The measured transverse momentum spread is consistent with what is expected from the transverse ground state energy. The anisotropy is hence a clear sign that we cool the atoms into a 2d gas, with the transverse degree of freedom frozen out. The in-plane width of the expanded cloud remains a good measure of temperature as the zero point motion is negligible in these dimensions. We however detect no signs of Bose condensation in the 2d traps; neither a bimodal distribution \cite{Krueger2007}, nor interference between the layers \cite{Hadzibabic2006} is observed. We therefore conclude that we create a set of thermal, rather than (quasi)condensed, 2d gases in the optical lattice on the atom chip. For the trap containing the maximal atom number ($N\sim 10^4$), an ideal gas would form a BEC ($N_c^{id}=3800$), but in the presence of interactions a Berezinskii-Kosterlitz-Thouless (BKT) type transition occurs at higher critical atom numbers. For the case of our layered traps, the dimensionless 2d coupling constant $\tilde{g}=0.46$ leads to a threefold increase of the BKT critical atom number with respect to the ideal gas BEC transition. The absence of signs of quantum degeneracy in our experiment confirms previous observations of BKT transitions in less strongly ($\tilde{g}=0.13$) interacting 2d gases \cite{Hadzibabic2008}.

In the second experiment we demonstrate coherent manipulation of atoms in the optical lattice by studying Bloch oscillations \cite{Bloch1929,BenDahan1996}. We start with a BEC created on the atom chip. We then switch off the wire trap, and 500 $\mu$s later switch on a weak optical lattice (too weak to hold the atoms against gravity).  After an adjustable interaction time, during which the Bloch oscillations occur, we switch off the optical lattice, and observe the atoms after a fixed total expansion time of 15 ms. 

\begin{figure}[t]\center
   \includegraphics[width=\columnwidth]{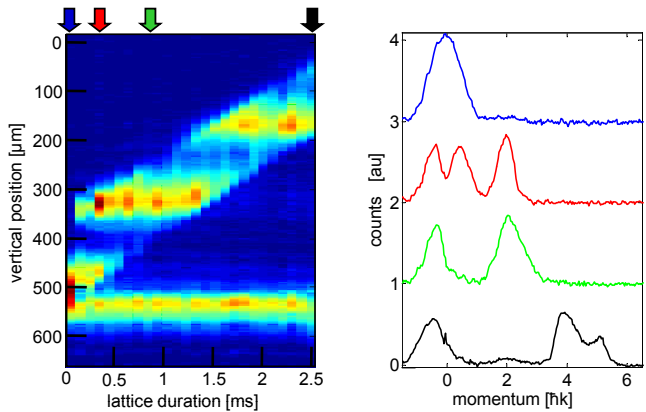}
   %\vspace{3cm}
   \caption{(\emph{color on-line}) Observation of Bloch oscillations of a BEC accelerated by gravity.
   (\emph{left}) Series of time-of-flight images with increasing interaction time with the optical lattice. Each vertical line corresponds to a single absorption image, integrated along the horizontal direction. Atoms with high initial velocities never fulfill the Bragg condition (bottom). Slower atoms are accelerated and successively scattered upwards, performing Bloch oscillations. (\emph{right}) Momentum distribution at specific times, indicated by arrows of corresponding color in the left graph.}\label{fig:BlochOsz}
\end{figure}

For the data displayed in Fig.~\ref{fig:BlochOsz} the time before switching on the lattice was set such that a small fraction of the atoms was already too fast to fulfill the Bragg condition $\hbar k = m v$ and continues to fall in gravity. Their position after 15 ms time-of-flight corresponds to the one of free falling atoms without any lattice potential. Slower atoms ($v < \hbar k /m=5.9\,\mu$m/ms where $k=4\pi/\lambda$ is the grating vector of the
lattice) get accelerated until they fulfill the Bragg condition and are ``caught'' in the optical lattice, continuously performing Bloch oscillations. The first two successive Bloch oscillations can clearly be seen in Fig.~\ref{fig:BlochOsz}.

Observation of Bloch oscillations close to a surface opens up new possibilities for precision atom-surface interaction measurements such as probing the Casimir-Polder potential. Atom chip based microtraps allow to accurately control the starting position of the atoms. To suppress dispersion effects due to interactions fermions \cite{Roati2004} or non-interacting bosons \cite{Gustavsson2008,Fattori2008} can be used.

In conclusion we have demonstrated that an optical lattice can be integrated on an atom chip for both trapping and coherent manipulation of ultra-cold atomic ensembles. This combination of two powerful tool in atom optics paves the path for many new possibilities ranging from studies of low-dimensional quantum systems to single-site addressability in a quantum register, to precision measurements of atom-surface interactions and other short-range forces.

We thank S. Wildermuth, M. Andersson, and E. Haller for help in the experiment and S. Groth for fabricating the atom chip. We acknowledge financial support from the Austrian Science Fund FWF and the European
Union (IP-SCALA, CHIMONO).

\end{document}